\documentclass[apj]{emulateapj}
\usepackage[colorlinks = true, linkcolor = blue, urlcolor  = blue, citecolor = blue, anchorcolor = blue]{hyperref}

\usepackage{multirow}
\usepackage{amsmath}
\usepackage{float}
\usepackage{multirow}

\usepackage{comment}
\usepackage[export]{adjustbox}
\newcommand{\na}{    {\it New Astron.}}
\newcommand{\comflu}{    {\it Computers \& Fluids}}

\shorttitle{Quasi-biennial oscillations in the temporal variation of chromospheric macrospicules}
\shortauthors{Kiss \& Erd\'{e}lyi, 2018}
\usepackage{comment}

\begin{document}

\title{On Quasi-biennial oscillations in chromospheric macrospicules and their potential relation to global solar magnetic field}

\author{T. S. Kiss \altaffilmark{1,2*} \& R. Erd\'elyi\altaffilmark{1,3}}
\thanks{\altaffilmark{*}e-mail: tskiss1@sheffield.ac.uk}
\affil{\altaffilmark{1}Solar Physics and Space Plasmas Research Centre (SP2RC), School of Mathematics and Statistics, University of Sheffield,\\ Hicks Building, Hounsfield Road, Sheffield S3 7RH, UK\\
\altaffilmark{2}Department of Physics, University of Debrecen, Egyetem t\'{e}r 1, Debrecen H-4032, Hungary\\
\altaffilmark{3}Department of Astronomy, E\"{o}tv\"{o}s Lor\'{a}nd University, Budapest, P\'{a}zm\'{a}ny P\'{e}ter s\'{e}t\'{a}ny 1/A, H-1117, Hungary}

\begin{abstract}
This study aims to provide further evidence for the potential influence of the global solar magnetic field on localised chromospheric jets, the macrospicules (MS). To find a connection between the long-term variation of properties of MS and other solar activity proxies, including e.g. the temporal variation of the frequency shift of solar global oscillations, sunspot area, etc., a database overarching seven years of observations was built up. This database contains 362 MS, based on observations at the 30.4 nm of the \textit{Atmospheric Imaging Assembly (AIA)} on-board the \textit{Solar Dynamics Observatory (SDO)}. Three of the five investigated physical properties of MS show a clear long-term temporal variation after smoothing the raw data. Wavelet analysis of the temporal variation of maximum length, maximum area and average velocity is carried out. The results reveal a strong pattern of periodicities at around 2-year (also referred to as Quasi-Biennial Oscillations -- QBOs). Comparison to solar activity proxies, that also possess the properties of QBOs, provides some interesting features: the minima and maxima of QBOs of MS properties occur at around the same epoch as the minima and maxima of these activity proxies. For most of the time span investigated, the oscillations are out-of-phase. This out-of-phase behaviour was also corroborated by a cross-correlation analysis. These results suggest that the physical processes, that generate and drive the long-term evolution of the global solar activity proxies, may be coupled to the short-term local physical processes driving the macrospicules, and, therefore modulate the properties of local dynamics.
\end{abstract}

\section{Introduction}
A wide variety of oscillations are already found in various layers of the Sun. Perhaps the first documented and maybe best-known periodicity is the nearly 11-year oscillation of the number of sunspots. Now, it is widely accepted that this variation is driven by the evolution of the global magnetic field of the Sun \citep{hathaway2015}. Since its strong influence on the structure of the solar corona, the understanding of the generation and evolution of the large-scale (global) solar magnetic field is one of the keys for a better understanding of our star. Studies, however, unveil another oscillation that also occur at large scale, that also seem to work alongside with the 11-year solar cyclic oscillation: the Quasi-Biennial Oscillations (QBOs). \cite{belmont1966} reported a pioneering analysis about the detection of a nearly two-year period oscillation. These authors found strong oscillations, with 19 months of period time, in the chromospheric 10.7 cm radioflux detected between February 1947 and December 1964 (Solar Cycles 18-19).

\begin{table*}[ht]
	\centering
	\caption{Dates and errors for extrema of the three MS properties}
	\begin{tabular}{|lll|lll|}
		\multicolumn{3}{c}{\textbf{Maximums}} & \multicolumn{3}{c}{\textbf{Minimums}}\\
		\textbf{No.} &  \textbf{Date} & \textbf{Error} & \textbf{No.} & \textbf{Date} & \textbf{Error}\\
		I. & 24.06.2011 & $\pm$ 28 days & II. & 15.07.2012 & $\pm$ 35 days\\
		III. & 24.05.2013 & $\pm$ 84 days & IV. & 15.03.2014 & $\pm$ 56 days\\
		V. & 01.06.2015 & $\pm$ 49 days & VI. & 07.04.2016 & $\pm$ 37 days 
	\end{tabular}
	\label{tab01}
\end{table*}
Since the discovery of \cite{belmont1966}, a good number of QBOs present in various solar structures have been either modelled theoretically or identified observationally. Hereinafter, we refer to oscillations as QBO, that have a period between 1.5 -- 3 years. 

On the theory side, \cite{zaqarashvili2010} investigated the stability of magnetic Rossby waves present in the solar tachocline. A shallow water approximation magnetohydrodynamic (MHD) code was used to model magnetic Rossby waves with periods around 2 years. The authors claim that periodic magnetic flux emergence may be generated this way, and this could be the root of QBOs. 

\cite{beaudoin2016} carried out a kinematic mean-field $\alpha^2 \Omega$ dynamo simulation employing an MHD generalization of the EULAG code \citep{prusa2008} with the aim to model the long- and short-term oscillations observed and briefly outlined above. This code works with two dynamo layers: one is placed at the bottom of the convective zone and generates the 11-year cycle, meanwhile the other dynamo operates near to the solar surface and is able to drive QBOs. 

Let us now briefly recall the type of QBOs that have already been reported. On the observation side, periodicities with a two-year period present underneath the solar surface, modulating the global $p$-mode oscillations, were somewhat wider reported. Global acoustic oscillations are trapped in cavities below the solar surface. These oscillations can be affected by the varying properties of the plasma environment they propagate in (e.g. temperature, structure, mean molecular weight). Observations show that a given mode of global solar $p$-mode oscillations reach their maximal frequencies during the solar maxima \citep{woodard1985}. To investigate the possible connection between the $p$-modes and the solar cycle, Sun-as-a-star (unresolved) Doppler velocity observations were carried out by the Birmingham Solar-Oscillations Network (BiSON; \citealt{chaplin1996}). Detailed analysis of the $p$-mode frequency shifts revealed a significant QBO period during the last three solar cycles, after filtering out the dominant 11-year oscillations \citep{broomhall2009}. This variation of the $p$-modes is in-phase with similar oscillations of other solar proxies such as the 10.7 cm radioflux and the Mg II core-to-wing ratio between 1977 and 2009. \cite{fletcher2010} divided the BiSON data into "low-frequency" (1.88 -- 2.77 mHz) and "high-frequency" (2.82 -- 3.71 mHz) sub-categories. Both datasets show the signature of QBOs.

Closer to the solar surface, i.e. in the photosphere, the detection of the majority of QBOs are connected to sunspots. \citet{akioka1987} studied several sunspot properties (such as sunspot area, sunspot number and their maximum throughout one solar rotation) of the late section of Solar Cycle 20 and the entire Solar Cycle 21 between 1969 and 1986. Power spectrum analysis of these data series shows a dominant oscillatory peak at around 17 months. \cite{benevolenskaya1998b, benevolenskaya1998a} claimed a possible existence of a double solar magnetic cycle based on the evolution of magnetic field patterns through Solar Cycle 21 and 22. In their view, the total magnetic oscillatory system is built up from two different components: a low frequency ($\approx 11$ years) and a high frequency ($\approx 2$ years) component. \cite{penza2006} studied the behaviour of three photospheric lines and reconstructed the cyclic variation of the full-disk line depths. Oscillation pattern was identified in both full-disk and center-disk data. These data can be decomposed into two dominant oscillations: one is the well-known 11-year cycle and another has periods of 2.8 years. The variation of both sunspot number and area ratio between the northern and the southern solar hemispheres is also possessing a QBO pattern \citep{badalyan2011, elek2018}. The temporal variation of the area of sunspot groups and the locii of solar flares within the active longitude also provide QBO signatures \citep{gyenge2013,gyenge2014, gyenge2016}.

\cite{kiss2017a} has recently built up a database of containing information on several physical parameters of chromospheric macrospicules (MS). A strong fluctuation in time with a period at around 2 years was found in the cross-correlation of macrospicule features \citep{kiss2017b}. This is a very intriguing result, as the properties of the basically short-lived (i.e. couple of minutes) macrospicules seem to show a long-term modulation of the order of years in a statistical sense. 

Even higher up in the solar atmosphere, in the corona, the long-term variation of the green coronal emission line at 530.3 nm indicates the dual presence of the 11-year and the QBO oscillations. Studies about the comparison of different QBOs at different layers or depth of the Sun were also carried out (\citealt{kane2005, broomhall2015}). Finally, closer to our planet, the geomagnetic activity $aa$-index provides a clear QBO pattern between 1844 and 2002 \citep{mursula2003}.

In this paper, we investigated the long-time variation of some key observed properties of macrospicules and compare their temporal variations to those of other solar activity proxies. The aim of this paper is to find evidence towards supporting the conjectured connection between the long-term variations of the physical properties of MS and that of the underlying magnetic field that may drive these local jets. In Section~\ref{database}, we give a detailed description of the observations and the obtained data. In Section~\ref{results}, we summarize our results and, finally, provide a brief discussion with conclusions in Section~\ref{discussion}.
\begin{figure*}
	\centering
	\includegraphics[width=\textwidth]{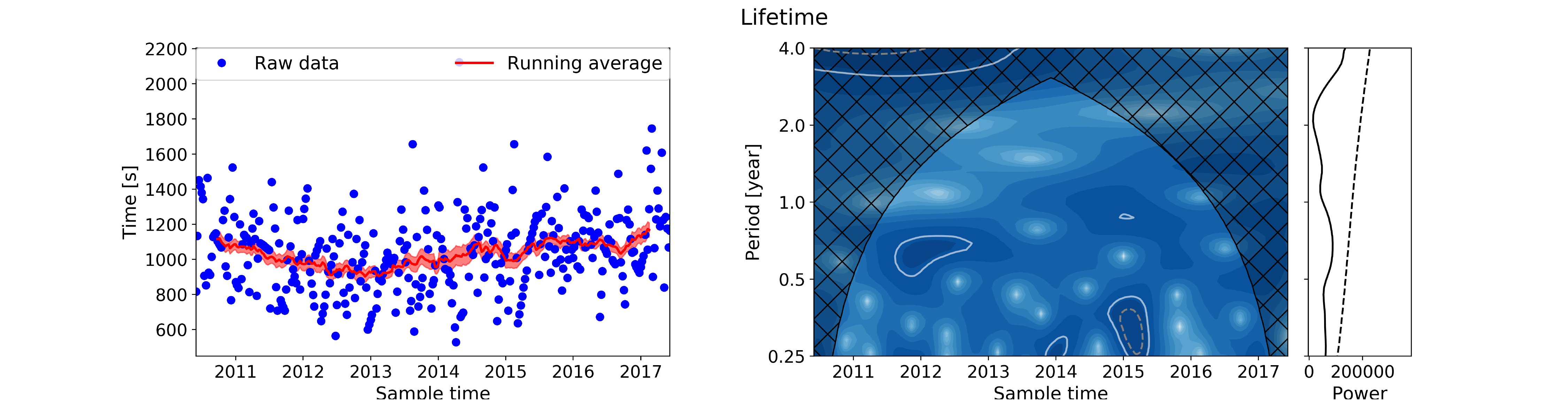}
	\includegraphics[width=\textwidth]{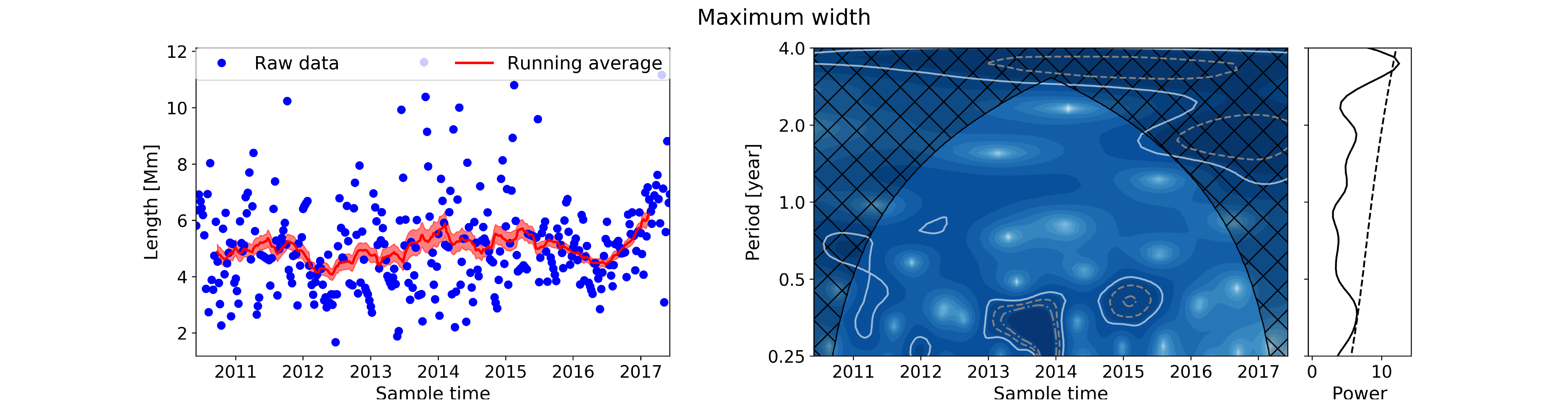}
	\includegraphics[width=\textwidth]{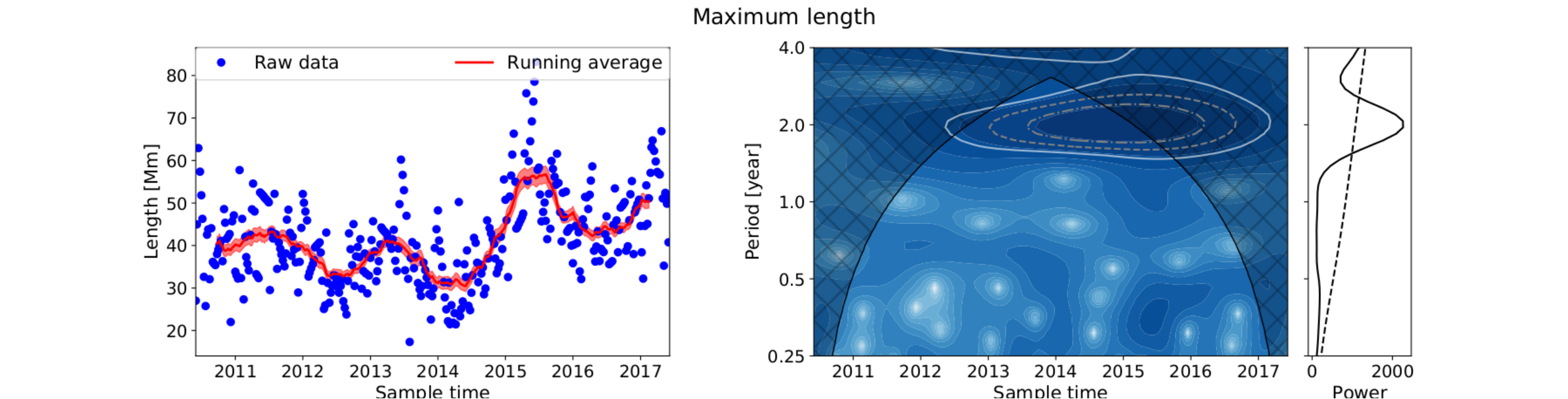}
	\includegraphics[width=\textwidth]{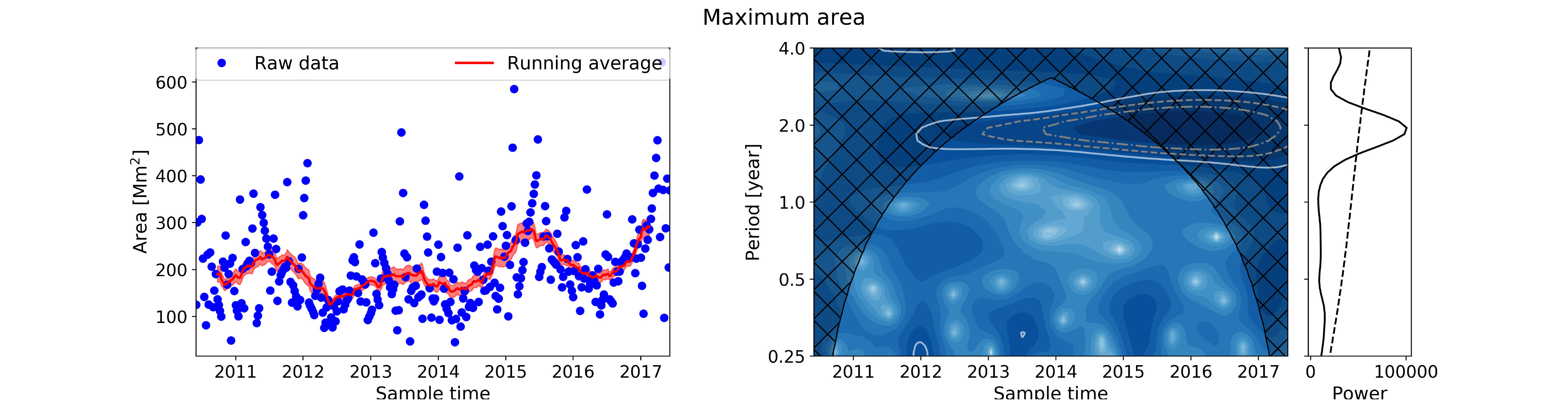}
	\includegraphics[width=\textwidth]{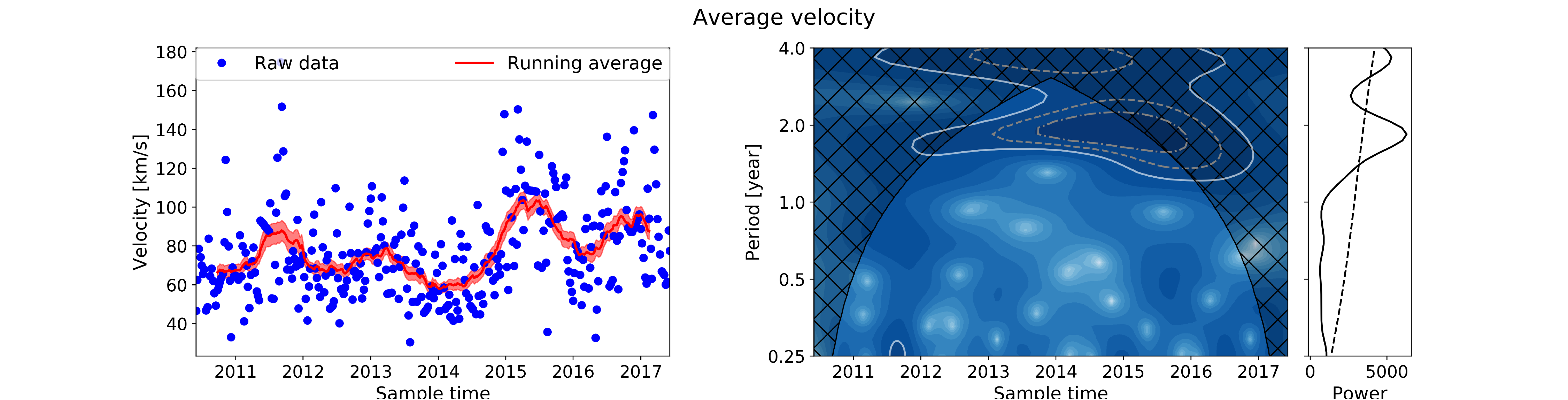}
	\caption{In the left column in each panel, raw data of the properties of MS are marked with blue dots and their running-average is highlighted with red dots with errorbar. In the right column, wavelet analysis of the row data of the five investigated physical properties of MS are shown. Solid, dashed and dotted contour lines represent the $1\sigma, 2\sigma$ and $3\sigma$ deviations from the average. The COI is bounded with a black line and filled with hatches. Attached to each wavelet on their right, the global wavelet power is seen.}
	\label{fig01}
\end{figure*}
\section{Database and methodology} \label{database}
\subsection{Macrospicules}
Macrospicules are formed likely in the chromosphere and have the potential to play an important role in the transfer of momentum and energy from the lower solar atmospheric regions to the transition region and the low corona. Several interesting features of MS have been found and reported since their first discovery \citep{bohlin1975}. These jets are multi-thermal \citep{pike1997}, rotating phenomena \citep{pike1998}, which can be observed both on-limb and off-limb \citep{scullion2010} with density about $10^{-10} \text{ cm}^{-3}$ \citep{parenti2002}. Their formation energy has been estimated \citep{bennett2015} and a possible connection to the active longitude of sunspots has been revealed \citep{gyenge2015}.

To gain relevant information about the long-term variation of the physical properties of MS, a multi-year long database is needed. The raw data are provided by the \textit{Atmospheric Imaging Assembly (AIA)} \citep{lemen2012} on-board \textit{Solar Dynamics Observatory (SDO)} \citep{pesnell2012} in the 30.4 channel. Identifying MS from the large variety of different chromospheric phenomena is a crucial point for this research, therefore firm criteria are necessary for defining what an MS may be. \cite{kiss2017a} discussed the range of such criteria in more details, the applied date selection method and how these jets are characterised as tetragons.

Accomplishment of the process of MS identification resulted in a catalogue of 363 MS between June 2010 and July 2017. The jets are categorized based on the solar environment around them at the solar limb: MS inside coronal hole and Quiet Sun regions are named as Coronal Hole MS [CH-MS] and Quiet Sun MS [QS-MS], respectively. The numbers of CH-MS and QS-MS are found approximately to be equal: 184 and 169. A third, so-called coronal hole boundary class was also constructed in order to account for MS with uncertainty on their more exact locii, but their number is much smaller (9). Therefore, these latter jets are not taken into account for our further study. 363 MS were found on 334 observing days, indicating $1.08$ MS to appear in the 2-hour time interval of a given day of observing.

For the tetragon approximation, four points were fitted to each MS. The geometric shape of tetragon grants a decent measure of the actual length, width and area for each \textit{SDO/AIA} during the entire lifetime of an MS. For our study, during the evolution of an individual MS, the maxima of these properties were recorded (e.g., maximum length, width and area, respectively) in addition to lifetime (the timespan between the first and the last \textit{SDO/AIA} observation of a given jet). The temporal resolution of \textit{SDO/AIA} at the wavelength of 30.4 nm is 12 s, therefore the actual velocity of the jet can be estimated by determining the length difference between two observations. The actual velocity of some individual MS may have outlier values, which are interpreted as errors. Therefore, an average velocity for each MS is used in the analysis instead of their maximum velocity.

Additional occasional interpolation is applied to the data of the maximum length, maximum width, maximum area, average velocity and lifetime series, because the temporal resolution is not perfectly homogeneous. The average number of MS occurrence is varying slightly on a given day of observation. Wavelet analysis is sensitive to the non-homogeneous nature of the input data. Consequently, a homogenizing method was constructed to overcome this problem. For the analysis, we need only a single data of each MS property per each epoch of observation. This approach addressed two issues: i) if two or more MS were observed on one observing block, the average of all of their physical properties is calculated, and, ii) if no MS is identified during the observing sequence then the value of each MS property on the given day is interpolated linearly from the closest existing values. We obtained the interpolated values by calculating the difference between neighbouring data points, divided the difference proportionately to the number of the missing dates, and, assigned a value by adding/extracting the fraction from the known data accordingly to the increasing/decreasing trend. During the built-up of the dataset, 330 dates were picked for investigation. On 140 days, only one MS was observed, therefore, these dates remained unaffected by the homogenisation process. In the case of 99 days, more than one MS were identified. On the remaining 91 dates, no MS observation was registered. Note, the homogenization will have no effect on the data at periods of the order of magnitude of QBOs. To study the temporal variation of oscillatory patterns of the homogenized dataset of MS properties, wavelet analysis was used.

\subsection{Complementary solar activity proxies}
In this section, we summarize proxies capturing long-term solar activity, what we will utilize in this study. 

1) The sunspot number is accounting for each sunspot and sunspot group on the visible surface of the Sun. This is one of the most important solar activity proxies for us today due their about 300 years of relatively continuous observations. All the data used here are provided on the website of the Sunspot Index and Long-term Solar Observation, Royal Observatory, Belgium (\url{http://sidc.be/silso/home}).

2) The sunspot area data indicates the total covered area of sunspot umbra and penumbra. These data are available at the late Debrecen Heliophysical Observatory (DHO) \url{http://fenyi.solarobs.csfk.mta.hu/en/databases/SDO/}). This catalogue is based on magnetograms of the \textit{Helioseismic and Magnetic Imager (HMI)} on-board the \textit{SDO} \citep{baranyi2016}.

3) The 10.7 cm radio flux is a bright emission of the upper chromosphere and the lower corona \citep{tapping1987}. Sources of this emission are the gyroresonances from thermal plasma trapped in magnetic field and bremsstrahlung of thermal plasma over active regions \citep{kundu1965, kruger1979}. There is a widely known correlation between the 10.7 cm radioflux and the solar sunspot activity. This emission is observed and integrated over the disc and the data made available by the Dominion Radio Astrophysical Observatory, National Research Council of Canada (\url{http://www.spaceweather.gc.ca/solarflux/sx-en.php}).

4) With six base stations all around the Globe, the Birmingham Solar Oscillations Network (BiSON) measures the line-of-sight velocity of the solar surface \citep{chaplin1996} (\url{http://bison.ph.bham.ac.uk/}). The observations are disc-integrated, so-called Sun-as-a-star observations, which are sensitive to low-$\ell$ $p$-mode oscillations, where $\ell$ is the degree of an eigenmode. The helioseismic data are divided into 182.5-day time series and the frequency shift of each individual global acoustic mode is calculated for $0 \leq \ell \leq 2$ and $2400 \leq \nu \leq 3500$ $\mu$Hz. 

5) The wavelength 94 $\text{\AA}$ is an optically thin, EUV intensity line. This emission is generated by the Fe XVIII ions at around 6 MK. The 94 $\text{\AA}$ channel of \textit{SDO/AIA} is mapping these Fe XVIII ions in the solar corona in every 12 s with the cadence of 0.6 arcsec ($\approx$ 435 km).
\section{Results} \label{results}
\subsection{Analysis of macrospicules}
In their recent study, \cite{kiss2017a} investigated the temporal variation of the physical properties of MS obtained for the period between June 2010 and December 2015. No obvious trends were seen, initially, in the data themselves. Therefore, the histograms of the data were analysed. These histograms were constructed in log-normal distribution. The cross-correlations of MS physical properties (e.g. "Maximum length" vs "Average velocity") do show a strong oscillatory pattern. However, timespan of the database in \cite{kiss2017a} was not sufficiently long to obtain confident enough results about the nature of periodicities in these oscillations. In \cite{kiss2017b}, they extended the dataset with another year of observation. With this extension, they confirmed that there is evidence for QBOs in the cross-correlations of MS physical properties. These facts suggest examining further the temporal evolution of the raw data in more details, in order to find the origin of these QBO signatures. The work here is about mounting evidence towards this goal.

The database in comparison to \cite{kiss2017a} has now been extended by an additional year and a half, i.e. expanding the number of MS, with an additional sample of 62 (27 CH-MS and 35 QS-MS). In all the panels on the left-hand side of Figure~\ref{fig01}, the temporally homogenized data of each investigated physical parameters of MS is given. To highlight the trends in these plots, the signals are smoothed. The temporal length of the running average is around a month. This temporal scale is long enough to mask the short-scale noise-like variations of the data. On the other hand, the window is short enough to prevent the influence on long-scale oscillations. 

The distributions of "Lifetime" and "Maximum width" do not show a clear oscillatory pattern, however, a strong one is visible in the case of "Maximum length", "Average velocity" and "Maximum area". Highest values of the cycles occur during the summer of each odd years: 2011, 2013, 2015. This trend suggests a period of around two-year, but a proper signal processing would be required to gain more precise (and with confidence) information about the oscillatory properties of these signals.

For this reason, wavelet analysis was applied to the homogenized data. The 7-year length of the datasets is sufficiently long enough to provide a power peak with the expected 2-year period outside the Cone of Influence (CoI). The right panel of each row, in Figure~\ref{fig01}, shows the results of the wavelet analysis. "Lifetime" and "Maximum width" do not show power peaks at around 2-year. However, wavelet of the remaining three properties of MS provides a much more promising outcome. A significant peak appears at around 1.8 years outside the CoI, which is interpreted as signature of QBOs. These oscillations are dominant after 2012, hence some parts of these oscillation peaks are inside CoI. However, there are highly significant areas of QBOs outside of the CoI. Some power peaks with shorter periods are also visible in these figures, but they are not as significant as the QBOs.

The global wavelets are confirming the statements above (see e.g. the right-hand-side of the wavelet panels in Figure~\ref{fig01}). "Lifetime" does not seem to provide any measurable oscillation peaks, furthermore "Maximum width" shows a more complex global wavelet with several non-significant power peaks. "Maximum length", "Average velocity" and "Maximum area" have clear and significant peaks at about 2-year (outside CoI).

\subsection{Macrospicules and other solar activity proxies}
\cite{broomhall2015} compared the long-term variation of the global $p$-mode frequency shift to a few other solar activity proxies for a period between 1985 and 2014: namely, two indicators in the photosphere (i.e. sunspot number, sunspot area), two proxies in the solar atmosphere (i.e. 10.7 cm radioflux, 530.3 nm green coronal index) and two activity phenomena present farther away from the solar surface (interplanetary magnetic field, the galactic cosmic-ray intensity). After the removal of the dominant 11-year long cycle, QBOs were found in each dataset. However, the actual details of these high-frequency oscillations were not compared to each other in that work.
\begin{figure*}[t]
	\centering
	\includegraphics[width=0.95\textwidth]{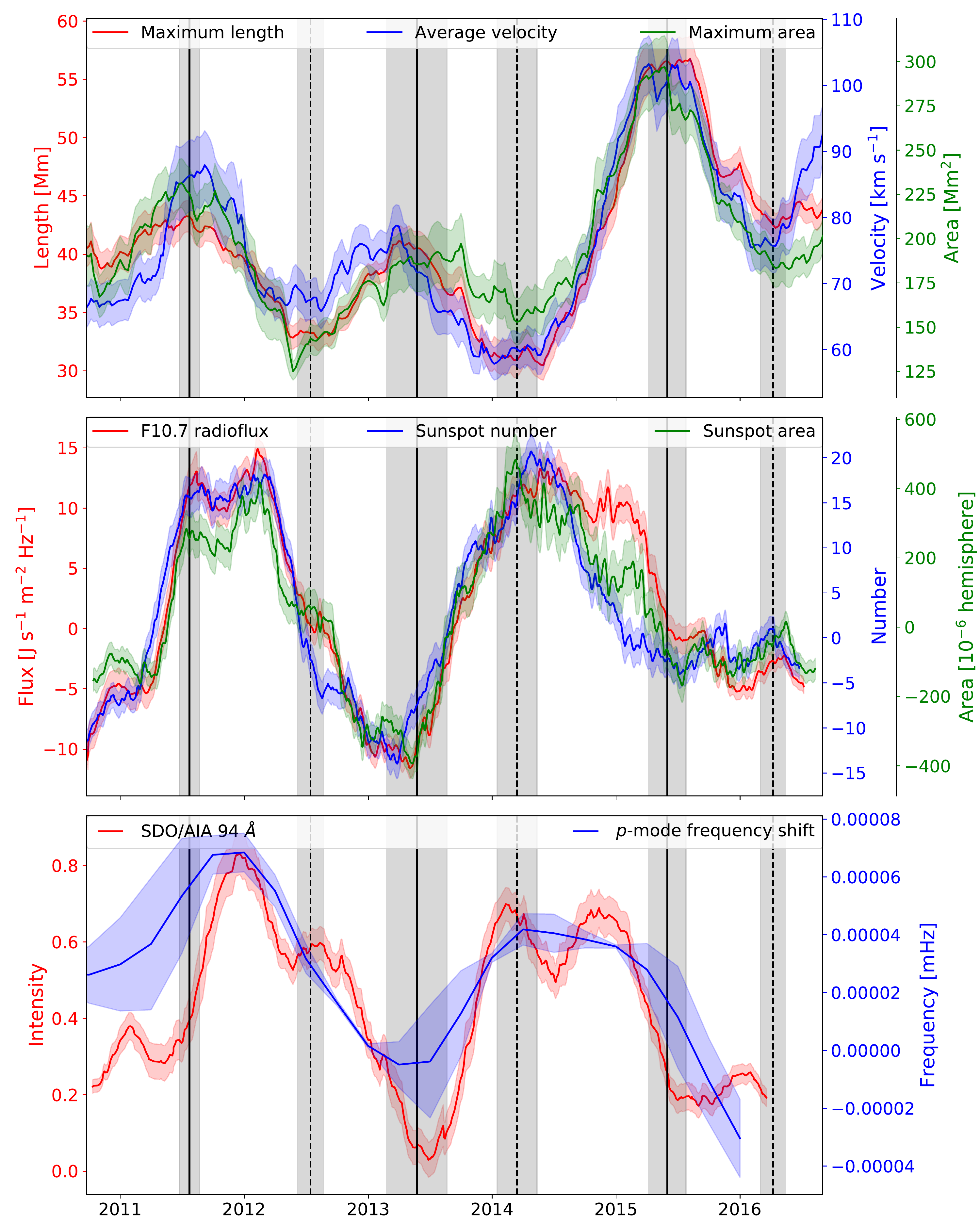}
	\caption{QBOs in various parameters of solar activity phenomena. The top panel contains three MS properties, which indicate QBOs: maximum length (red solid line), average velocity (blue solid line), maximum area (green solid line). In the middle panel, 10.7 cm radioflux (red solid line), sunspot number (blue solid line) and sunspot area (green solid line) are plotted. \textit{SDO/AIA} 9.4 nm intensity (red solid line) and the $p$-mode frequency shift (blue solid line) are showed. Coloured, partially transparent areas around each line indicates the error. Maximum and minimum dates of MS variables are represented with black vertical lines: solid vertical lines reveal the time of the maximums, dashed vertical lines illustrate the time of the minimums. Gray bars around the vertical lines demonstrate the error of the extrema.}
	\label{fig02}
\end{figure*}

\begin{figure*}[t]
	\centering
	\includegraphics[width=0.95\textwidth]{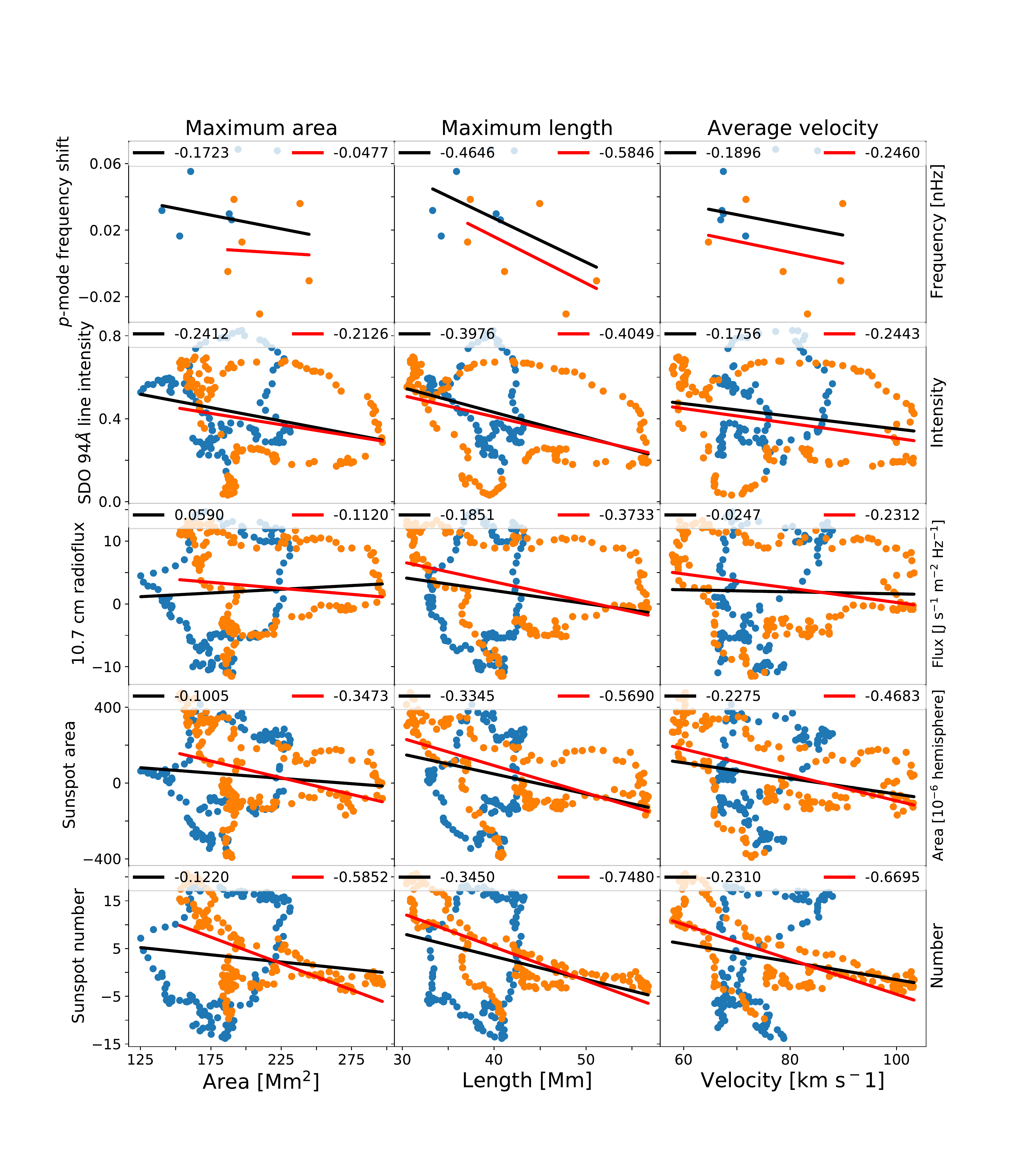}
	\caption{The cross-correlations between the three MS properties and the five solar activity proxies. In each panel, blue and orange data points represent together the entire time-span. The solid black line shows linear correlation between the datasets and the $r^2$ value can be found on the LHS of the legends. Orange points demonstrate the data after May of 2013. Their linear correlations are displayed by red lines with $r^2$ values on the RHS of the legends.}
	\label{cc}
\end{figure*}

Here, besides highlighting the high-frequency variation of the physical properties of MS, we also shed light on the temporal variations of different QBOs in comparison to each other. For this reason, we have also removed the 11-year cycle from the time series of 5 solar activity proxies (and labelled them with abbreviation in square brackets): the sunspot number [$n_{\text{SN}}$], area [$n_{\text{SA}}$], 10.7 cm radioflux [$\mathcal{F}_{\text{10.7}}$], average channel intensity of SDO/AIA 9.4 nm [$I_{\text{SDO}}$] and the $p$-mode frequency shift [$d_p$]. The residual data, after removing the solar cycle trend, are plotted in the bottom two panels of Figure~\ref{fig02}. The trend removal was executed in the following way: a smoothing was utilized with a large windowsize (nearly 2.5 years), which has removed shorter-scale fluctuations but preserved the 11-year cycle. The smoothed graph clearly shows the 11-year oscillation. Next, this smoothed graph is extracted from the original dataset and the residual is shown in the middle and bottom panels of Figure~\ref{fig02}. All the SDO channels present QBO patterns, however, the 9.4 nm channels provide the most similarities to other solar activity proxies. "Maximum length", "Maximum area" and "Average velocity" of MS, reveal QBOs, that are clearly visible in the top panel of Figure~\ref{fig02}. The duration of the MS dataset allows to carry out this study now between June 2010 and July 2017.  

In the next step, we compare the oscillatory signatures of the MS data to other solar proxies. Local extrema (minima and maxima) can be determined for each MS datasets. To find the epochs of all the MS extrema, the dates of the local extrema of the "Maximum length", "Maximum area" and "Average velocity" datasets were averaged (solid vertical lines for maxima and dashed vertical lines for minima in Figure~\ref{fig02}) and their standard deviations (gray vertical bars around the lines Figure~\ref{fig02}) were estimated for error analysis. Dates for all MS extrema are given in Table~\ref{tab01} (labelled with roman numbers I--VI). The dates of the extrema clearly indicated supporting evidence for the 2-year long oscillation, only the minima in 2014 may seem to be somewhat early.
 
In the following, we now compare these extrema with those of the other solar proxies:

\begin{itemize}
	\item[I:] this maximum is close to one of the maxima of $n_{\text{SN}}$-$n_{\text{SA}}$-$\mathcal{F}_{\text{10.7}}$, though, not matching them really well. With a good approximation, however, they are \textit{in-phase}. $I_{\text{SDO}}$ and $d_p$ are in their rising phase.
	\item[II:] this minimum occurs around at the time, when there is a small "step" in each of the other proxies.
	\item[III:] this maximum is matching the local minimums of all other proxies. They are clearly \textit{out-of-phase}.
	\item[IV:] this minimum happens at the same time, when there is a local maximum in the $n_{\text{SN}}$-$n_{\text{SA}}-d_p$ variations. In the case of $\mathcal{F}_{\text{10.7}}$ and $I_d$, IV also marks the local maximum, however, these variables have a smaller secondary maximum. Overall, they are all \textit{out-of-phase}.
	\item[V:] this maximum is similar to that of I. At this epoch, $n_{\text{SN}}$-$n_{\text{SA}}$ are around their local minimum, but the matching is not perfect. $\mathcal{F}_{\text{10.7}}$-$I_d$ is in its decaying phase from their secondary maximum, closing up to the minimum. Overall, they are all \textit{out-of-phase}. 
	\item[VI:] Close to the epoch of this minimum, a small-scale maxima is visible on the $n_{\text{SN}}$ - $n_{\text{SA}}$ - $\mathcal{F}_{\text{10.7}}$ - $I_{\text{SDO}}$ plots.

\end{itemize}

For the six extrema of the MS properties, four of them are in-- or out-of-phase with the other solar proxies. Only at II there is no extrema, but a "step" is still visible in all the other proxies. This may suggest a change in the underlying process: before this time, I is \textit{in-phase} with the other proxies and after this time, III is \textit{out-of-phase} with other proxies. 

To corroborate this \textit{out-of-phase} behaviour, we implemented a cross-correlation analysis between the three MS time-series and the five solar activity proxies as seen in Figure~\ref{cc}. To present a proof, anti-correlation should be found on the cross-correlation plots. Linear regression was utilized to analyse the correlation between every combination of them. By taking into account the entire time-span of the data-series, no strong anti-correlations were found (the average of $r^2$ values is -0.218). By considering the data points only after May 2013 (the epoch of III), the anti-correlation between the MS properties and the solar proxies is significantly stronger (the average of $r^2$ values is -0.3895). In 13 cases out of 15 cross-correlations, the absolute value of the correlation coefficients is larger. Figure~\ref{cc} clearly shows that the blue points (data values before May 2013) are robustly vertical, which weakens the correlation coefficient and, thus, the anti-correlation. This is a supporting evidence for the previously discussed \text{out-phase} nature.

However, for a more detailed investigation of these variation of extrema, longer data would be needed. Therefore, this analysis may be a promising motivation for investigation in the future.

\section{Discussion and conclusions} \label{discussion}
In our work, we focus on the investigation of the long-term variation of properties of chromospheric jets, the macrospicules. In addition to our previous analyses \citep{kiss2017a, kiss2017b}, 1.5 years of observation has now been added to the length of the data series, including data of further 62 jets. The full length of this dataset enables now to study more comfortably periodicities present in the property signals of MS at around the period of two-year, the so-called Quasi-Biennial Oscillations. To find such periodicities in the five time-series of MS properties ("Maximum length", "Maximum width", "Maximum area", "Average velocity", "Lifetime"), wavelet analysis was applied.
\begin{figure*}[t]
	\includegraphics[width=\textwidth]{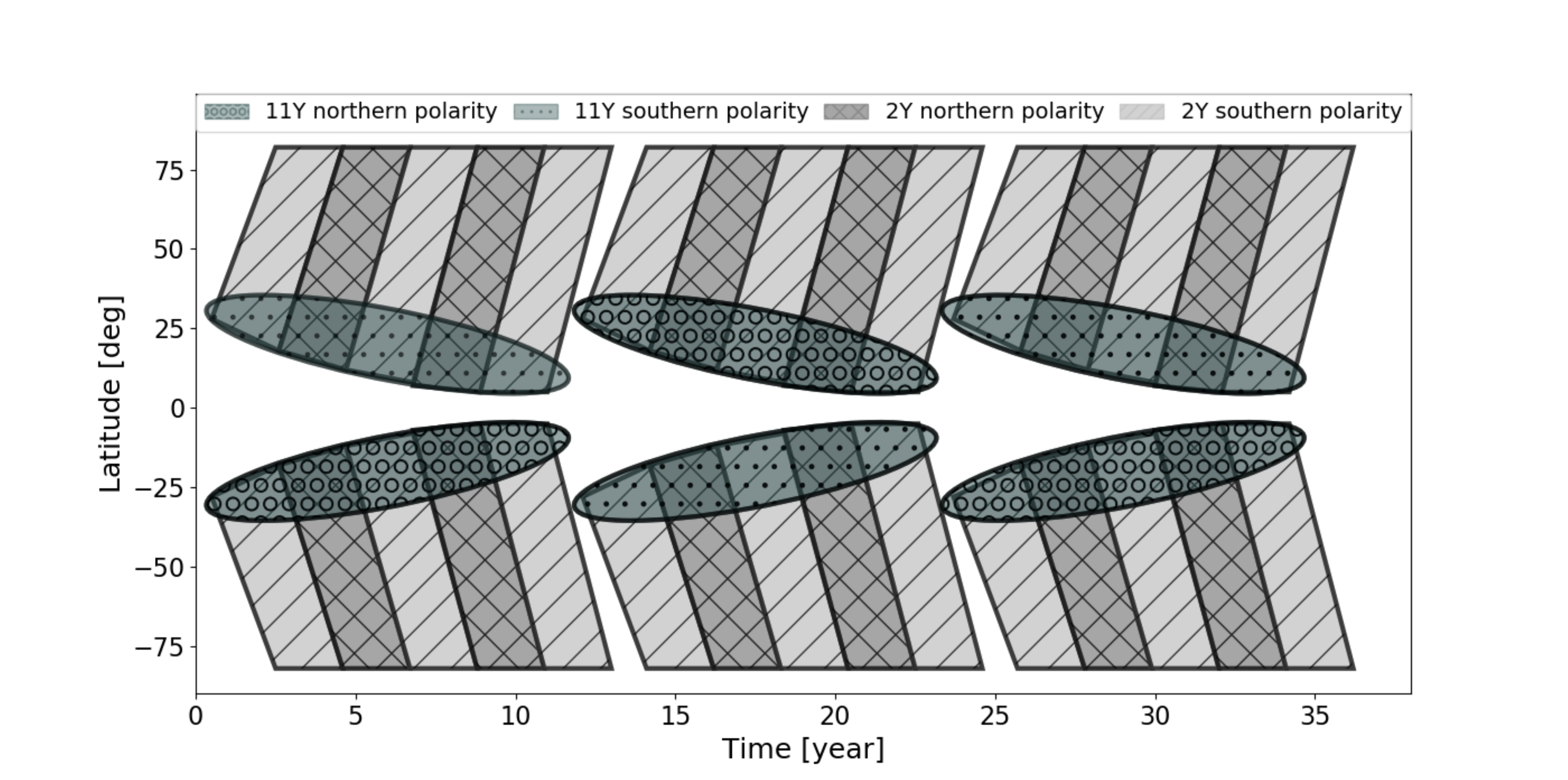}
	\caption{Schematic figure about our concept of the spatial distribution of the double solar magnetic cycle. Dark-grey ellipses around the solar equator represent the nearly 11-year long cycle (circled hatches are indicating regions where the polarity of the leading sunspots are northern, dotted hatches are showing the southern one). Between the "butterfly wings" and the poles, light-grey territories are demonstrating the cycles of QBOs. Correspondingly to the 11-year long oscillation, the striped and crossed hatches are distinguishing the cycles of the two dominate polarities.}
	\label{fig03}
\end{figure*}
Fundamentally, the five time-series of properties can be divided into two classes: the "Lifetime" and the "Maximum length" do now show any QBOs, however, the "Maximum length", "Maximum area" and "Average velocity" provide strong evidence, one outside of the CoI, mostly after 2012. 

A study comparing QBOs present in the properties of MS to the five additional solar activity proxies (namely, sunspot number, sunspot area, 10.7 cm radioflux, \textit{SDO/AIA} 9.4 nm intensity, $p$-mode frequency shifts) was carried out as well. For the period of June 2010 to July 2017, six extrema (local minimum or maximum) were identified in the QBOs of MS (labelled with I--VI) in Figure~\ref{fig02}. To have a deeper understanding of these QBOs, we compared directly the QBOs of MS to the QBOs of other solar activity proxies. All of the extrema indicate a local change of trend in the solar proxy oscillations. Five of them [I, III, IV, V, VI] are close to a local minimum or maximum of the solar proxy QBOs. Furthermore, III and IV are matching with high accuracy the local extrema of the solar proxy plot. 

Another interesting feature is that I is \textit{in-phase}, III-IV-V-VI are \textit{out-of-phase} with the activity proxies. The shift between the phases may take place during the epoch of II, where there is no extremum in the activity proxies, but a slight trend-breaking "step" is also visible. This \textit{out-of-phase} behaviour is also corroborated by a cross-correlation analysis. The linear regression shows a weak anti-correlation between the three MS properties and the five solar activity proxies during the studied time-span ($r^2 \sim -0.2$). However, the cross-correlation after May 2013 (the epoch of III) presents a stronger anti-correlation ($r^2 \sim -0.4$) as seen in Figure~\ref{cc}.

This analysis indicates a possible connection between solar proxies, hence the high-frequency oscillatory component of the magnetic field, and the locally driven macrospicules.

As a theoretical conjecture, we report our hypothesis about the evolution of the global magnetic field of the Sun. Previous studies suggest that the large-scale magnetic field is built up from two components: a low-frequency one with a nearly 11-year period and a high-frequency element, which produces the QBO oscillation. We suppose that, in each component, the polarity of the magnetic field is changing from one cycle to an other. Connecting this idea to the theory of the magnetic flux emergence, the likely physical background of the evolution of both solar activity proxies (\citealt{parker1955, zwaan1985}) and jets \cite{sterling2015} can be confirmed.  

Figure~\ref{fig03} illustrates the spatial distribution of both components in small-scale structures, like MS. The low-frequency components appear close to the solar equator and produce the widely-known butterfly diagram. As seen in Figure~\ref{fig02}, QBOs tend to show up both around the equator (in the temporal evolution of the solar activity proxies) and the poles (in the physical parameters of MS). Due to these facts, we assume that QBOs are forming bands between the poles and the equator. The QBOs are overlapping spatially with the 11-year cycle, which would support the double-faced nature of the solar activity proxies. However, the accurate behaviour of QBOs (e.g. their inclination to the 11-year oscillation) is still unknown, therefore, being a potential subject of further research and modelling. 

\section{Acknowledgement}
The authors thanks ESPRC (UK) for supporting this research. RE is grateful to STFC (UK), grant number ST/M000826/1, and The Royal Society. Some data were provided by William Chaplin, BiSON group, University of Birmingham, and guidance for $p$-mode frequency shift calculation are also acknowledged. The authors also thank Norbert Gyenge for the useful advice received.

\section{References}
\bibliographystyle{aasjournal}

\end{document}